\documentclass{emulateapj}

\usepackage{amssymb, amsmath, apjfonts, natbib, color}

\def\hv{{$h_3 - \overline{V}$}}
\def\vm{{$\overline{V}$}}

\def\degree{{\circ}}
\newdimen\digitwidth
\setbox1=\hbox{0}
\digitwidth=\wd1
\catcode`"=\active
\def"{\kern\digitwidth}

\slugcomment{Accepted for Publication in ApJ}

\begin{document}

\title{Shape of LOSVD\lowercase{s} in barred disks: Implications for future IFU surveys}

\author{Zhao-Yu Li\altaffilmark{1, 2, 3, 4}, Juntai Shen\altaffilmark{1, 2, 5}, Martin Bureau\altaffilmark{6}, Yingying Zhou\altaffilmark{1, 2}, Min Du\altaffilmark{7}, and Victor P. Debattista\altaffilmark{8, 9}}

\altaffiltext{1}{Key Laboratory for Research in Galaxies and Cosmology, Shanghai Astronomical Observatory, Chinese Academy of Sciences, 80 Nandan Road, Shanghai 200030, China}
\altaffiltext{2}{College of Astronomy and Space Sciences, University of Chinese Academy of Sciences, 19A Yuquan Road, Beijing 100049, China}
\altaffiltext{3}{Correspondence should be addressed to: jshen@shao.ac.cn; lizy@shao.ac.cn}
\altaffiltext{4}{LAMOST Fellow}
\altaffiltext{5}{Newton Advanced Fellow of the Royal Society}
\altaffiltext{6}{Sub-department of Astrophysics, Department of Physics, University of Oxford, Denys Wilkinson Building, Keble Road, Oxford OX1 3RH, UK}
\altaffiltext{7}{Kavli Institute for Astronomy and Astrophysics, Peking University,
Beijing 100871, China}
\altaffiltext{8}{Jeremiah Horrocks Institute, University of Central Lancashire, Preston, PR1 2HE, UK}
\altaffiltext{9}{Center for Theoretical Astrophysics and Cosmology, Institute for Computational Science, University of Z\"urich, Winterthurerstrasse 190, CH-8057 Z\"urich, Switzerland}

\begin{abstract}

The shape of LOSVDs (line-of-sight velocity distributions) carries important information about the internal dynamics of galaxies. The skewness of LOSVDs represents their asymmetric deviation from a Gaussian profile. Correlations between the skewness parameter ($h_3$) and the mean velocity (\vm) of a Gauss-Hermite series reflect the underlying stellar orbital configurations of different morphological components. Using two self-consistent $N$-body simulations of disk galaxies with different bar strengths, we investigate \hv\ correlations at different inclination angles. Similar to previous studies, we find anticorrelations in the disk area, and positive correlations in the bar area when viewed edge-on. However, at intermediate inclinations, the outer parts of bars exhibit anticorrelations, while the core areas dominated by the boxy/peanut-shaped (B/PS) bulges still maintain weak positive correlations. When viewed edge-on, particles in the foreground/background disk (the wing region) in the bar area constitute the main velocity peak, whereas the particles in the bar contribute to the high-velocity tail, generating the \hv\ correlation. If we remove the wing particles, the LOSVDs of the particles in the outer part of the bar only exhibit a low-velocity tail, resulting in a negative \hv\ correlation, whereas the core areas in the central region still show weakly positive correlations. We discuss implications for IFU observations on bars, and show that the variation of the \hv\ correlation in the disk galaxy may be used as a kinematic indicator of the bar and the B/PS bulge.

\end{abstract}


\section{INTRODUCTION}
\label{sec:intro}

Kinematic information is essential to understand disk secular evolution. It encapsulates the potential, angular momentum, and underlying stellar orbits of the disk and bar, if present. Measurement of the disk kinematics can reveal the disk formation history, the bar and spiral arms growths and evolutions, and allow to estimate the dynamical mass of the whole disk.

Integral-field unit (IFU) spectroscopic observations of nearby disk galaxies provide 2D spatially-resolved spectral information, whereby important kinematic properties can be measured. They are a powerful tool to investigate bar kinematics \citep[e.g.][]{cappel_etal_07, krajno_etal_11}. IFU surveys such as ATLAS$^{\rm 3D}$ \citep{cappel_etal_11}, CALIFA \citep{sanche_etal_12, garcia_etal_15}, SLUGGS \citep{brodie_etal_12, brodie_etal_14}, SAMI \citep{croom_etal_12, bryant_etal_15}, MaNGA \citep{bundy_etal_15}, and MUSE \citep{bacon_etal_10} have led to significant progress in our understanding of disk galaxy formation and evolution.

The shape of line-of-sight velocity distributions (LOSVDs) can be described by Gauss-Hermite series. Key kinematic information includes the mean velocity (\vm), velocity dispersion ($\sigma_{\rm los}$), and the third and fourth Gauss-Hermite coefficients $h_3$ and $h_4$, describing the asymmetric (``skewness'') and symmetric (``kurtosis'') deviations from a Gaussian profile, respectively \citep{gerhar_93, van_fra_93, bender_etal_94}. Positive $h_3$ indicates a high-velocity tail, and negative $h_3$ a low-velocity tail. For $h_4$, a positive value indicates a sharp central peak, and a negative value results from a flat-top profile. Correlations between $h_3$ and the mean line-of-sight velocity \vm\ reflect the underlying stellar orbits.

Commonly seen in disk galaxies, bars play important roles in their secular evolution \citep[e.g.][]{kor_ken_04}.  \citet{bur_ath_05} used $N$-body simulations to confirm the negative \hv\ correlation (i.e. anticorrelation) observed in edge-on disks and the positive correlation observed in bars, as reported in previous long-slit observations of edge-on galaxies \citep[e.g.][]{fisher_etal_97, chu_bur_04} and the theoretical orbital analysis by \citet{bur_ath_99}. They suggested that LOSVDs with a high-velocity tail (positive \hv\ correlation) may be tracers of bars. \citet{debatt_etal_05} also suggested that when viewed face-on, boxy/peanut-shaped (B/PS) bulges tend to show $h_4$ values in the inner regions that are smaller than elsewhere \citep[also see][]{ian_ath_15}. In doubly barred disks, \citet{du_etal_16} found peaks in the line-of-sight (LOS) velocity dispersion $\sigma_{\rm los}$ near the inner bars, and \hv\ anticorrelations in the inner bars for certain orientations. These Gauss-Hermite coefficients relations are important indices that can be used to understand bar kinematics and evolution when compared to IFU observations.

\begin{figure*}
\plotone{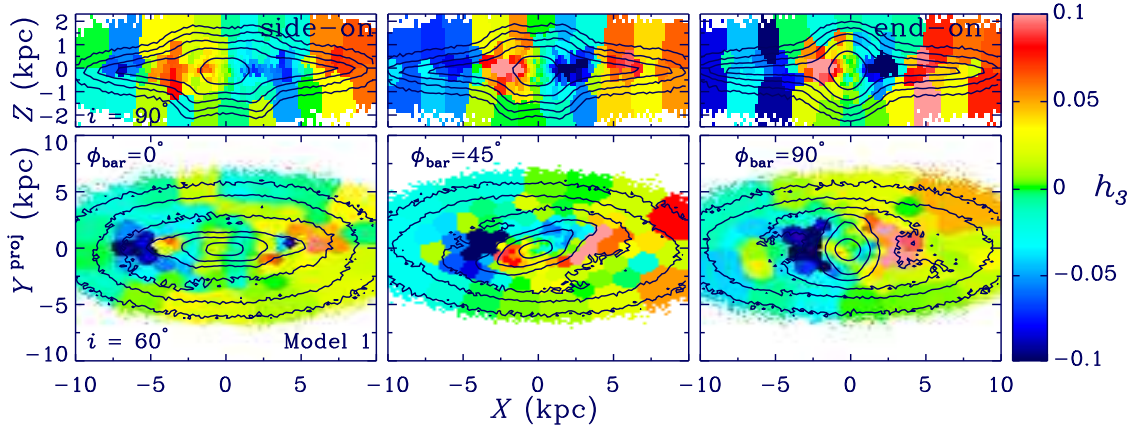}
\caption{$h_3$ maps of Model~1 viewed edge-on ($i = 90^\degree$, top row) and partially inclined ($i = 60^\degree$, bottom row), with different bar viewing angles ($\phi_{\rm bar} = 0^\degree, 45^\degree$, and $90^\degree$ in the first, second, and third column, respectively). The disks rotate clockwise as seen from $Z > 0$, i.e.\ positive (receding) at $X < 0$ and vice versa. The projected density contours are overlaid on the maps. As the inclination decreases, the \hv\ correlations in the outer part of the bar area change from positive to negative, while the core area dominated by the B/PS bulge still maintains weak positive \hv\ correlations. The anticorrelations persist in the disks (with smaller $h_3$ amplitudes) as the inclination decreases.}
\epsscale{1.}
\label{fig:map_60_90_m1}
\end{figure*}

\begin{figure*}
\plotone{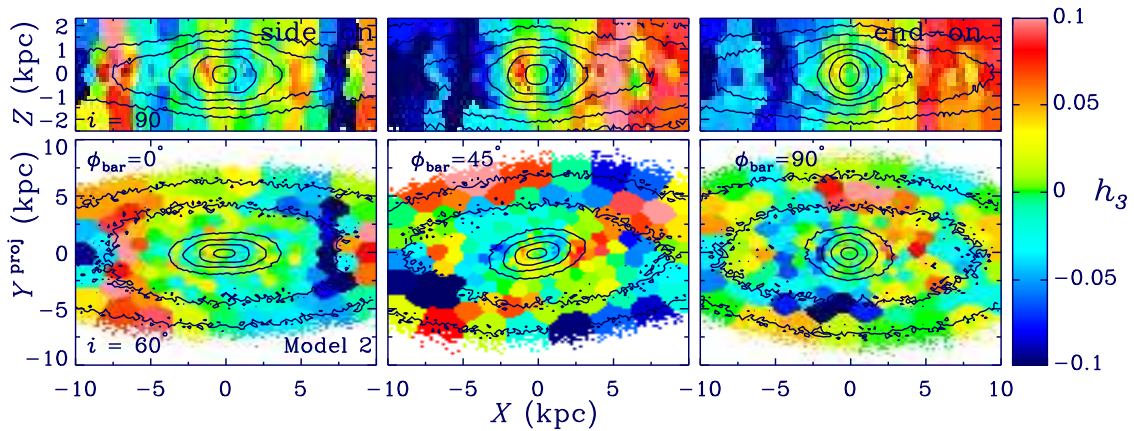}
\caption{$h_3$ maps of Model~2 viewed edge-on ($i = 90^\degree$, top row) and partially inclined ($i = 60^\degree$, bottom row), with different bar viewing angles. The disks rotate clockwise as seen from $Z > 0$. The projected density contours are overlaid on the maps. Model~2 shows similar $h_3$ pattern as Model~1. Note that in the first column (Model~2 with $\phi_{\rm bar} = 0^\degree$), there is a positive \hv\ correlation in the outer disk along the bar major axis ($|X| \sim 8$ kpc), which may be due to the Outer Lindblad Resonance of the bar.}
\epsscale{1.}
\label{fig:map_60_90_m2}
\end{figure*}

To understand the \hv\ (anti-)correlations, and the reasons behind those correlations, we carry out a study based on two self-consistent $N$-body simulations of disk galaxies with different bar strengths. We also investigate the inclination angle (and bar viewing angle) dependence of key kinematic features, especially in the B/PS bulge regions.

The paper is organized as follows. Section 2 describes the two simulations. The results and corresponding discussion are presented in Sections 3 and 4, respectively. Key results are summarized in Section 5.

\begin{figure*}
\plotone{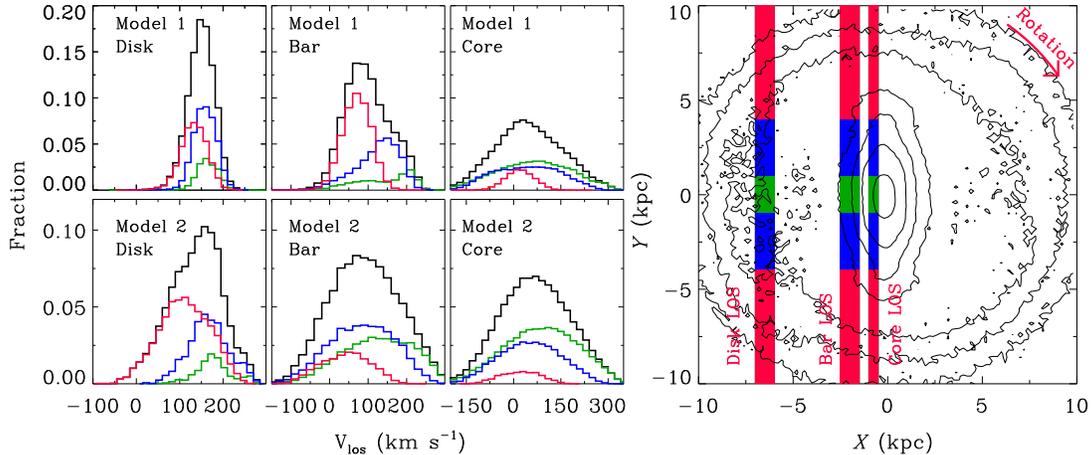}
\caption{LOSVDs in the disk and bar areas of the two models viewed edge-on ($i = 90^\degree$) with an end-on bar ($\phi_{\rm bar} = 90^\degree$). The black histograms are the combined LOSVDs of all particles in the three areas. The green, blue, and red histograms represent the tangent, intermediate, and wing particles (foreground and background), respectively. The right panel sketches the three lines-of-sight in the disk, bar, and core areas, and the choice of the tangent, intermediate, and wing regions on top of the iso-density contours of Model~1. The disks rotate clockwise. In the disk areas, the wing particles lead to the low-velocity tails of the LOSVDs, producing characteristic negative \hv\ correlations. In the bar area, the tangent particles lead to high-velocity tails and thus positive \hv\ correlations.}
\label{fig:vhist_90}
\end{figure*}

\section{SIMULATIONS}
\label{sec:model}

Two disk galaxy $N$-body simulations with different bar amplitudes are analyzed here. Face-on and edge-on projections of the two models are shown in the top and middle panels of Fig.~1 in \citet{li_she_15}. Compared to Model 2, the bar in Model 1 is longer and stronger, that has experienced higher buckling instability, resulting in a more prominent B/PS bulge \citep{com_san_81, raha_etal_91}. Initially, the two models are featureless exponential disks. In Model~1, two million disk particles evolve in a live dark matter halo, consisting of 2.5 million particles with a compressed King profile ($\Psi(0) / \sigma^2 = 3$ and $r_{\rm c} = 10 R_{\rm d}$; see \citealt{sel_mcg_05} for details of the adiabatic compression).  Model~2 was shown to reproduce well the photometric and kinematic properties of the Galactic Bulge in \citet{shen_etal_10}. It consists of one million disk particles rotating in a rigid dark matter halo potential. Bars in both models grow quickly to form an inner B/PS bulge. In previous studies, the two models (especially Model~2) have been used extensively to understand the structure and kinematic properties of the Galactic bulge and disk \citep[e.g.][]{shen_etal_10, li_she_12, li_etal_14, molloy_etal_15a, molloy_etal_15b, nataf_etal_15, qin_etal_15, li_she_15}.

To understand the inclination dependence of key kinematic properties, the two models are projected with moderate and edge-on inclination angles ($i = 60^\degree, 90^\degree$) and different bar viewing angles ($\phi_{\rm bar} = 0^\degree, 45^\degree, 90^\degree$).\footnote{Close to face-on ($i \lesssim 30^\degree$), LOS velocities are too small to show clear $h_3$ pattern.} $\phi_{\rm bar}$ is the angle between the major axis of the bar and the major axis of the inclined disk ($\phi_{\rm bar} = 0^\degree$ is thus a side-on bar and $\phi_{\rm bar} = 90^\degree$ an end-on bar). To calculate \vm, $\sigma_{\rm los}$, $h_3$, and $h_4$, we fit the Gauss-Hermite series to the LOSVDs up to the fourth order.\footnote{We also fit the Gauss-Hermite series up to the eighth order. The fourth-order and eighth-order fit yield consistent results, with the fifth and higher order coefficients being negligible.} $h_3$ maps of the edge-on disks ($i = 90^\degree$) and moderately inclined disks ($i = 60^\degree$) in Models~1 and~2 are shown in Figs.~\ref{fig:map_60_90_m1} and~\ref{fig:map_60_90_m2}, respectively. We used the Voronoi binning method of \citet{cap_cop_03} (with $S/N = 50$). The disks in the two models rotate clockwise as seen from $Z > 0$, i.e.\ positive (receding) at $X < 0$ and vice versa. 

\section{Results}
\label{sec:results}

\begin{figure*}
\plotone{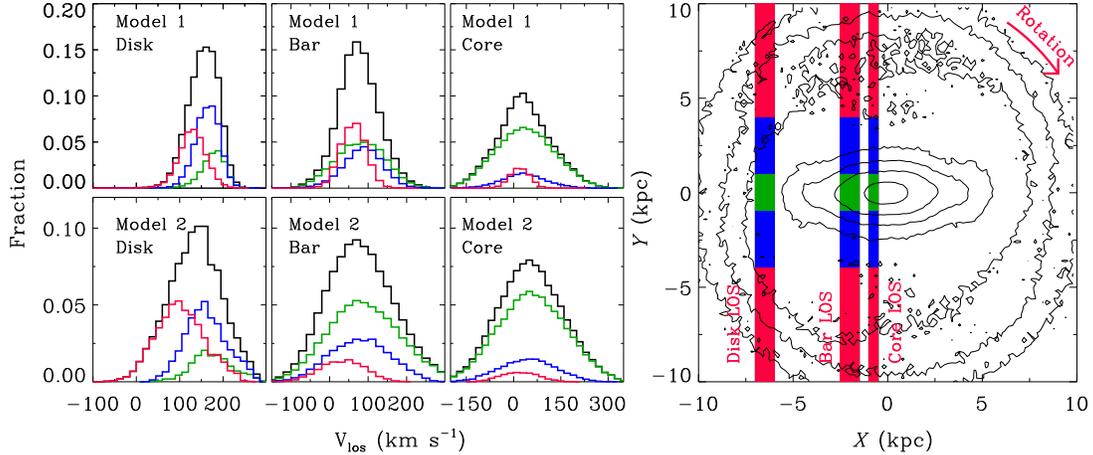}
\caption{LOSVDs in the disk and bar areas of the two models viewed edge-on ($i = 90^\degree$) with a side-on bar ($\phi_{\rm bar} = 0^\degree$). The layout of this figure is the same with Fig.~\ref{fig:vhist_90}. LOSVDs in the bar regions are more symmetric than Fig.~\ref{fig:vhist_90}.}
\label{fig:vhist_90_sideon}
\end{figure*}

\begin{figure*}
\plotone{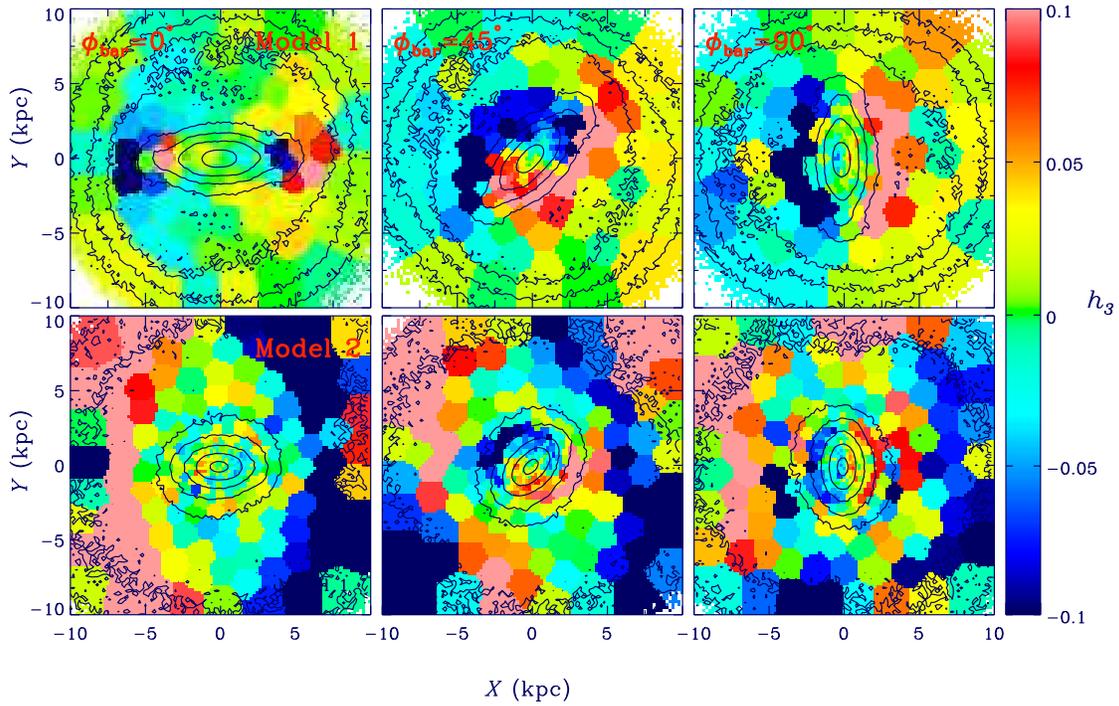}
\caption{$h_3$ maps of Model~1 (top row) and Model~2 (bottom row) viewed face-on with different bar angle (from left to right, $\phi_{\rm bar} = 0^\degree,\ 45^\degree$, and $90^\degree$). $h_3$ here is derived from $V_{Y}$, which represents the LOS velocity in the edge-on view with the observer in the $X-Y$ plane and the LOS perpendicular to the $X$-axis. The disks rotates clockwise. In the outer part of the bar, $h_3$ displays anticorrelation with $\overline{V_Y}$, consistent with the analysis in Figs.~\ref{fig:vhist_90} and ~\ref{fig:vhist_90_sideon}.}
\label{fig:maps_faceon_h3}
\end{figure*}

For Model~1, in the edge-on view (top row in Fig.~\ref{fig:map_60_90_m1}), inside the bar area ($|X| \lesssim 4$ kpc), $h_3$ and \vm\ exhibit positive correlations, with larger $h_3$ values in the end-on cases ($\phi_{\rm bar} = 90^\degree$; third column) than the side-on cases ($\phi_{\rm bar} = 0^\degree$; first column). In the outer disks ($|X| \gtrsim 4$ kpc), $h_3$ generally anti-correlates with \vm, as expected. These results are consistent with previous studies \citep{bur_ath_05} and the kinematics of the Milky Way bar/bulge \citep{zhou_etal_17}. 

For moderately inclined disks ($i = 60^\degree$, bottom row of Fig.~\ref{fig:map_60_90_m1}), inside the outer parts of the bars, the \hv\ correlations change drastically from positive to negative, whereas the core areas dominated by the B/PS bulges (within about half the bar length) still show weak positive correlations. This is most significant in the second and third columns for Model~1 ($\phi_{\rm bar} = 45^\degree$ and $90^\degree$). This phenomenon was first noticed in \citet{bur_ath_05} and confirmed in \citet{ian_ath_15}. They found that the \hv\ correlation in the bar region decreases as $i$ decreases, and even becomes anticorrelation for $i \lesssim 80^\degree$. This is consistent with our results, except the core areas, which maintain positive \hv\ correlations. From the top left panel in Fig.~\ref{fig:map_60_90_m1}, it seems that the iso-density contours of Model~1 in the edge-on view with a side-on bar are not symmetric with respect to the mid-plane. The buckling event may not be completely finished. We have performed similar analysis in earlier snapshots of this simulation. The main results are unchanged.

For Model~2, as shown in Fig.~\ref{fig:map_60_90_m2}, it exhibits consistent $h_3$ pattern with Model~1 with smaller $h_3$ amplitude and weaker \hv\ correlation. Note that in the first column of Fig.~\ref{fig:map_60_90_m2}, the disk area at $|X| \sim 8$ kpc shows a positive \hv\ correlation along the major axis of the bar. This feature may be related to the Outer Lindblad Resonance (OLR) of the bar in Model~2. According to \citet{bin_tre_08}, we estimate the location of OLR of the two models by comparing the bar pattern speed ($\Omega_{\rm b}$) and the radial profiles of $\Omega_{\rm b} + \kappa / 2$. The location of OLR is $\sim$ 14 kpc for Model~1 ($\Omega_{\rm b} \approx 20\ \rm km\ s^{-1}\ kpc^{-1}$), and $\sim$ 8 kpc for Model~2 ($\Omega_{\rm b} \approx 40\ \rm km\ s^{-1}\ kpc^{-1}$). The bar rotates slower in Model 1, thus resulting in a larger OLR radius and the absence of this feature in the disk of Model 1.

We also examined $h_4$ maps of the two models. In the face-on views, the B/PS bulges show $h_4$ values that are smaller than elsewhere, in agreement with \citet{debatt_etal_05} and \citet{ian_ath_15}.

\section{DISCUSSION}
\label{sec:discussion}

\subsection{LOSVDs in disk and bar areas}

The most direct way to understand the \hv\ correlations is dissecting the LOSVDs themselves. We thus select typical fields in the disk and bar areas. For the edge-on and moderately inclined viewing angles, the selected disk areas are at $-7 < X < -6$ kpc, while the selected bar areas are at $-2.5 < X < -1.5$ kpc. The selected core areas are at $-1 < X < -0.5$ kpc. All three areas are restricted with $|Z| < 0.5$ kpc. 

\subsubsection{Edge-on disks ($i = 90^\degree$)}

Theoretically, for an axisymmetric edge-on disk, the main peak of the LOSVD at any given position is contributed by particles at the tangent point of nearly circular orbits, with a low-velocity tail from the projected stellar orbits away from (both in front and beyond) the tangent point \citep{bur_ath_05}. This generates the usual \hv\ anticorrelation. 

The LOSVDs of the disk and bar areas in the edge-on view with an end-on bar are shown in Fig.~\ref{fig:vhist_90}. As shown in the right panel, along each LOS, the particles are separated into three subsamples according to the distance from the tangent point, representing a tangent region (green, $< 1$ kpc), an intermediate region (blue, $1 - 4$ kpc), and a wing region (red, $> 4$ kpc). The corresponding LOSVDs of the disk areas in the two models are shown in the first column with the same color scheme. As expected, the green histograms peak at the highest velocities, $\sim 170$ km s$^{-1}$, while the blue and red histograms peak at lower velocities,  $\sim$ 160 and 100 km s$^{-1}$, respectively. This agrees with the theoretical expectations that the tangent points lead to the highest velocities, while the wing regions lead to low velocities. For the total LOSVDs, the low-velocity tails are mainly contributed by the red histograms. The combined effect is the usual \hv\ anticorrelation. 

The three-dimensional orbital configuration of bars has been extensively studied in the literature \citep{pfenni_84, pfe_fri_91, skokos_etal_02a, skokos_etal_02b, patsis_etal_02, patsis_etal_03}. Inside the bar, the main orbit families of $N$-body simulations can be regarded as three-dimensional generalizations of the main two-dimensional orbit family, i.e. the $x_1$ family corresponding to orbits elongated parallel to the bar \citep{bur_ath_05}. The $x_1$ orbits lead to higher LOS velocities than the outer orbits in the end-on bar view, and much lower velocities in the side-on bar view \citep{bur_ath_99}. As the viewing angle to the bar approaches end-on, the velocity distribution from the elongated orbits in the bar shifts towards velocities higher than those of circular orbits. Considering the low-velocity contributions from quasi-circular (foreground and background) projected disk orbits surrounding the bar, the combined effect is a high-velocity tail, i.e.\ an unusual positive \hv\ correlation.

For our two models, the velocity distributions in the bar field are shown in the second column of Fig.~\ref{fig:vhist_90}. Clearly, the green histograms show much larger peak velocities ($\sim$ 200 km s$^{-1}$) than those of the blue histograms (intermediate regions). The red histograms (wing regions) show the lowest peak velocities ($\sim$ 50 km s$^{-1}$). This is fully consistent with the theoretical expectations of the $x_1$ orbit family. However, in the bar areas, both the green and blue histograms show low-velocity tails, i.e.\ large negative $h_3$ values (when considered independently). This is more significant in Model~1, with a strongly buckled bar ($h_3^{\rm tangent} = -0.462$). The LOSVDs inside the bar areas show low-velocity tails and large dispersion. The $x_1$ orbits seem unlikely to contribute to such low-velocity tails. Other orbital families are needed to explain this feature.


As shown in the third column of Fig.~\ref{fig:vhist_90}, the core areas have large velocity dispersions. The green histograms' peak velocities are slightly higher than those of the blue histograms. The LOSVDs in tangent and intermediate regions are quite symmetric, with small $h_3$ values. The total LOSVDs show weak positive \hv\ correlations.

We also study the LOSVDs in the bar and disk areas in the edge-on disk with a side-on bar. The results are shown in Fig.~\ref{fig:vhist_90_sideon}. In the bar area, the LOSVDs become much broader than the end-on case, resulting in a much weaker $h_3$ amplitude and \hv\ correlation.

Fig.~\ref{fig:maps_faceon_h3} shows the face-on view of the $h_3$ maps of $V_{Y}$ for the two models in different bar angles. $V_{Y}$ actually represents the LOS velocity in the edge-on view with the observer in the $X-Y$ plane and the corresponding LOS perpendicular to the $X$-axis. This provides a clear visualization of the spatial distribution of $h_3$ values in different regions of the galaxy. The disks rotate clockwise. In the bar region, the outer part exhibits clear anticorrelation with \vm. This is consistent with Figs.~\ref{fig:vhist_90} and~\ref{fig:vhist_90_sideon}. The disk of Model~2 at $|R| \sim 8$ kpc shows positive \hv\ correlation, probably due to the influence of OLR.

\subsubsection{Moderately inclined disks ($i = 60^\degree$)}

\begin{figure}
\epsscale{1.2}
\plotone{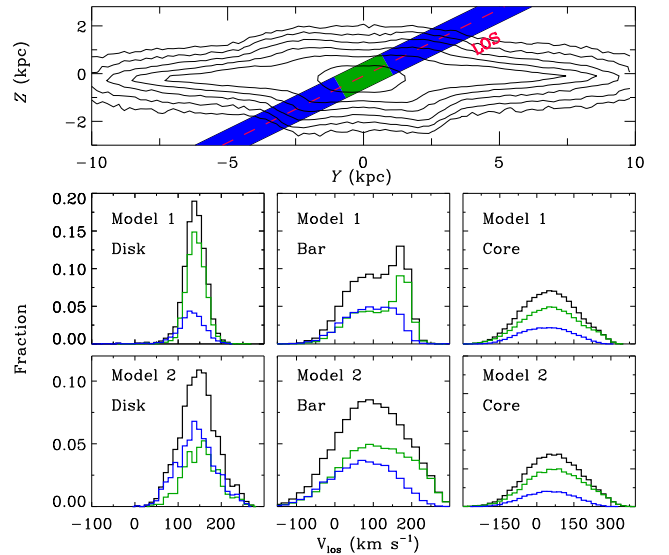}
\caption{LOSVDs in the disk and bar areas of the two models viewed at $60^\degree$ inclination with an end-on bar ($\phi_{\rm bar} = 90^\degree$). Top panel shows the LOS through the iso-density contours of Model~1 in the edge-on view with a side-on bar, with green and blue regions representing the tangent and intermediate regions, respectively. The black histograms are the combined LOSVDs of all particles in the three areas. The green and blue histograms represent the tangent and intermediate region particles, respectively. The disk areas now have very small $h_3$ values, i.e.\ symmetric LOSVDs, due to the lack of particles in the wing region. The LOSVDs in the bar areas show prominent low-velocity tails, and thus the unusual \hv\ anticorrelations. In the core area, the total LOSVDs maintain positive \hv\ correlation.}
\epsscale{1.}
\label{fig:vhist_60}
\end{figure}

At slightly smaller inclination angles ($i = 60^\degree$), \hv\ correlations change dramatically. As shown in Figs.~\ref{fig:map_60_90_m1} and~\ref{fig:map_60_90_m2}, $h_3$ values in the disk area decrease at smaller inclinations, while inside the bar areas, especially in the end-on cases, the \hv\ correlations change from positive to negative in the outer parts of the bars. The core areas maintain weak positive correlations.

For lines-of-sight through the disks at $i = 60^\degree$ with an end-on bar, as shown in the top panel of Fig.~\ref{fig:vhist_60}, the depths along the lines-of-sight (8 kpc) are roughly twice the vertical thickness of the disks (4 kpc).\footnote{The top panel shows the iso-density contours in the $Y-Z$ plane. Therefore the bar is in side-on view. The selected disk, bar and core areas have different $X$ ranges, but the same $Y$ and $Z$ ranges. The red dashed line represents the projected LOS of the disk, bar and core areas.} For planar orbits, a $60^\degree$ inclination will cause a 13\% reduction of \vm. The velocity distributions will thus shift to slightly lower velocities and become narrower. We again divide the particles into two groups, i.e.\ the tangent region (green) and an intermediate region (blue). Due to the smaller depths along the lines-of-sight, we cannot define a wing region $\gtrsim 4$ kpc away from the tangent point, as before.

For the disk areas, as shown in the left column of Fig.~\ref{fig:vhist_60}, the green histograms peak at $\sim 160$ km s$^{-1}$, the blue ones at $\sim 140$ km s$^{-1}$. The $h_3$ values are very small for the combined histograms, due to the lack of low-velocity contributions from the wing regions as in the edge-on cases.

In the outer part of the bar, in the middle column of Fig.~\ref{fig:vhist_60}, both the green and blue histograms show clearly skewed distributions with low-velocity tails. Here, in the moderately inclined cases, the lines-of-sight do not go through the outer disk areas. Without the low LOS velocity contributions from the outer disks, the bars themselves display the unusual \hv\ anticorrelations. The velocity distribution of the tangent region (green histogram) in Model~1 seems to be composed of two components, i.e.\ a narrow peak $\sim$ 170 $\rm km\ s^{-1}$ and a broad peak $\sim$ 50 $\rm km\ s^{-1}$. Similar feature can also be seen in the green histogram (tangent region) in the top middle panel of Fig.~\ref{fig:vhist_90} for the bar area. The high velocity narrow peak is probably related to the $x_1$ orbits, while the broad low velocity peak may be due to other orbital families in the bar.

In the core area, as shown in the right column of Fig.~\ref{fig:vhist_60}, the LOSVDs of the tangent and intermediate regions have large velocity dispersions and small $h_3$ values. The weak positive \hv\ correlations at $60^\degree$ inclination in the core area are consistent with the edge-on cases, where the LOSVDs are dominated by the tangent and intermediate regions.

We also tried considering only the bar particles in the simulations. Regardless of the inclination angles, there are positive and negative \hv\ correlations in the core and outer bar areas, respectively, consistent with the previous argument.

In this work, we found similar phenomena as \citet{ian_ath_15}, but we strive to understand the reason behind this. Along each LOS, by dissecting the particles into tangent, intermediate and wing regions, we investigate the shape of the LOSVDs causing the correlation. For the bar area, the dependence of the \hv\ correlation on the inclination angle is also explained with the combination of the LOSVDs from the tangent, intermediate and/or wing regions.

\subsection{Comparison with IFU observations of nearby galaxies}

Direct comparison between our simulation predictions and the IFU observations is not straightforward. $h_3$ derived from the observed spectra is sensitive to dust extinction correction and has large fluctuations \citep{seidel_etal_15}. Moreover, bars are often accompanied with nuclear disks, nuclear bars or pseudobulges, that may affect the \hv\ correlation in the core region. 

Our pure bar models at moderate inclinations ($i \sim 60^\degree$) predict a positive \hv\ correlation in the core and an anticorrelation in the outer part of the bar. In edge-on view, the simulations exhibit a positive \hv\ correlation in the bar region. According to previous simulations \citep{bur_ath_05, ian_ath_15}, the \hv\ signature is strongest with relatively large inclination angles  ($i \gtrsim 40^\degree$) and bar viewing angles ($\phi_{\rm bar} \gtrsim 40^\degree$). \citet{seidel_etal_15} studied 2D kinematics of 16 barred galaxies observed with SAURON IFU. Unfortunately, only 2 galaxies have relatively large inclination angles ($\sim 60^\degree$) and large $\phi_{\rm bar}$, i.e. NGC~2543 and NGC~5350. For the two galaxies, we do see \hv\ anticorrelations in the outer part of the bar, consistent with our predictions for the bar kinematics at moderate inclination angles.  Similar \hv\ anticorrelation in the bar region was reported in \citet{saburo_etal_17} with long-slit observations of UGC~1344. 

If the iso-density contours are more peanut-like, with weak $h_3$ values, then the bar is probably aligned with the projected major axis of the disk, i.e.\ $\phi_{\rm bar} = 0^\degree$. If the iso-density contours are more spheroidal with strong positive correlations between $h_3$ and \vm, $\phi_{\rm bar}$ is close to $90^\degree$. Recently, \citet{opitsc_etal_17} mapped the kinematics of the M31 bulge region, and found a positive \hv\ correlation. Considering the relatively large inclination angle of M31 (close to edge-on), this positive correlation and the lack of clear B/PS isophotes probably imply the existence of an end-on bar in M31, also consistent with our predictions. 

Our simulations do not include nuclear substructures in the central region of the bar. For real galaxies, additional substructures, e.g. nuclear disk, secondary bars, or pseudobulges \citep{kor_ken_04}, could overwhelm the weak \hv\ correlation predicted by our models. In fact, \citet{seidel_etal_15} detected \hv\ anticorrelations in the core regions in half of their sample. The detected anticorrelation probably hints for a kinematically decoupled substructure, since a pure bar model predicts weak positive correlations in the core region. In the future, we will make more tests with more sophisticated simulations including the later formation of nuclear substructures \citep[e.g.][]{cole_etal_14}.

\subsection{Bars and B/PS bulges identification}

In edge-on galaxies, the disks generally show \hv\ anticorrelations, while the bars display positive correlations. Therefore, based on the area showing a positive \hv\ correlation in long-slit or IFU observations, the bar existence may be revealed. The bar roughly corresponds to the region with positive \hv\ correlation. In addition, the $h_3$ amplitude depends on the viewing angle of the bar. An end-on bar usually shows larger $h_3$ values than a side-on bar. 


In a moderately inclined disk, the bar can be directly measured from the image at small $\phi_{\rm bar}$. When $\phi_{\rm bar}$ is close to $90^\degree$, a bar may be difficult to identify. In this case, negative \hv\ correlations in the inner region of the disk can help to confirm the existence of a bar.

Recent observational studies have identified B/PS bulges in local disk galaxies \citep{erw_deb_17, li_etal_17}. From our results, for a barred galaxy with B/PS bulge in moderate inclinations, without the presence of nuclear substructures, the core area could be identified by the central positive \hv\ correlation, with the outer part of the bar showing negative \hv\ correlation. This feature is different from results of the pure bar models without B/PS bulges. \citet{ian_ath_15} investigated the individual contribution of bars and B/PS bulges on the observed kinematics by comparing a simulation in pre- and post-B/PS formation. They found quite significant differences in velocity dispersion, $h_3$ and $h_4$ maps, with B/PS bulges showing strong $h_3$ and $h_4$ features off the kinematic-major axis. The in-plane values are also boosted with B/PS bulge. At smaller inclination angles, the simulations without B/PS bulges show negative \hv\ correlation in the bar and the core regions. We performed similar analysis by analyzing the snapshots of our models with pre- and post-B/PS formation. The results are consistent with \citet{ian_ath_15}.

\section{SUMMARY}
\label{sec:summary}

We use two $N$-body simulations of disk galaxies with different bar strengths to investigate their disk kinematics and dependence on inclination. For the disks viewed edge-on, we confirm the negative and positive \hv\ correlations in the disk and bar areas, respectively. The $h_3$ amplitude is larger in bars viewed end-on than side-on. At $60^\degree$ inclination, the $h_3$ amplitude in the disk areas is smaller, while in the bar areas, the \hv\ correlations change from positive to negative in the outer parts of bars, and remain weakly positive in the core area dominated by the B/PS bulge. 

To understand the origin of the \hv\ correlation, we dissect the LOSVDs in the bar and disk areas at different inclination angles. In the edge-on views, for the disk areas, the tangent region of the underlying quasi-circular orbits along the LOS leads to the highest velocities, while the regions far from the tangent point mainly lead to a significant low-velocity contribution. The combined distribution thus shows an \hv\ anticorrelation. In the bar areas, the bar particles lead to velocities even higher than those of circular orbits, changing the combined LOSVD to one with a high-velocity tail, thus resulting in a positive \hv\ correlation. These results are consistent with theoretical expectations. However, for bar particles only, the LOSVDs in the outer part of the bar show significant low-velocity tails, which seems unlikely to be contributed by $x_1$ orbits.

At a $60^\degree$ inclination, the depths along the lines-of-sight decrease from $\sim$ 20 kpc to $\sim$ 8 kpc. All velocities decrease by $\sim$ 13\% due to the projection effect. Because of the smaller depths along the lines-of-sight, there is no contribution from low-velocity disk particles surrounding the bars. Therefore the total LOSVDs are mainly contributed to by the tangent and intermediate regions. In the disk areas, the LOSVDs are fairly symmetric with very small $h_3$, while in the bar areas, the LOSVDs of the tangent and intermediate regions show negative \hv\ correlations in the outer parts of the bars, and weak positive correlations in the core areas. This results in the observed contrasting behavior with respect to the edge-on cases, and is confirmed by a test using the bar particles, where both the edge-on and moderately inclined disks show anticorrelations between $h_3$ and \vm\ in the outer parts of the bars, and weak positive correlations in the core areas.

We also compare with IFU observations and find our predictions roughly consistent with IFU and long-slit observations of nearby galaxies in the bar region, suggesting that the \hv\ correlation could be a good indicator for the bar identification. For a disk viewed edge-on, the bar can be associated with the area showing a positive \hv\ correlation. In moderately inclined disks, $h_3$ and \vm\ are anti-correlated in the outer parts of the bars. In long-slit and IFU observations of moderately inclined disks, the anticorrelations seen in the bar areas are thus fully consistent with bar kinematics; there is no need for an additional disk to explain the anticorrelations. 

From our results, we can see that the core region dominated by the B/PS bulge shows positive \hv\ correlation at moderately inclination angle, whereas the outer parts of the bars show negative \hv\ correlation. This feature is absent for simulations without B/PS bulges, that only show weak \hv\ anticorrelations in the bar region. This drastic feature in the bar area can be used in B/PS bulge identification.


We thank the anonymous referee for constructive suggestions that helped to improve the paper. The research presented here is partially supported by the 973 Program of China under grant no.\ 2014CB845700, by the National Natural Science Foundation of China under grant nos.\ 11773052, 11333003, 11322326, and 11403072, and by a China-Chile joint grant from CASSACA. ZYL is supported by the Youth Innovation Promotion Association, Chinese Academy of Sciences. His LAMOST Fellowship is supported by Special Funding for Advanced Users, budgeted and administrated by Center for Astronomical Mega-Science, Chinese Academy of Sciences (CAMS). JS acknowledges support from an {\it Newton Advanced Fellowship} awarded by the Royal Society and the Newton Fund, and from the CAS/SAFEA International Partnership Program for Creative Research Teams. MD is supported by the ``National Postdoctoral Program for Innovative Talents'' grant from the China Postdoctoral Science Foundation. VPD is supported by STFC consolidated grant \#ST/M000877/1 and acknowledges the personal support of George Lake, and of the Pauli Center for Theoretical Studies, which is supported by the Swiss National Science Foundation (SNF), the University of Z\"urich, and ETH Z\"urich during a sabbatical visit in 2017. This work made use of the facilities of the Center for High Performance Computing at Shanghai Astronomical Observatory. 


\end{document}